# Extension of the B3LYP – Dispersion-Correcting Potential Approach to the Accurate Treatment of both Inter- and Intramolecular Interactions.

Gino A. DiLabio[1,2] * Mohammad Koleini,[1,3] and Edmanuel Torres[1,4]

[1]National Institute for Nanotechnology, National Research Council of Canada, 11421 Saskatchewan Drive, Edmonton, Alberta, Canada T6G 2M9; [2]Department of Physics, University of Alberta, Edmonton Alberta, Canada T6G 2E1; [3]Department of Chemical and Materials Engineering, University of Alberta, Edmonton Alberta, Canada T6G 2V4; [4]Faculty of Basic Sciences, Universidad Tecnológica de Bolívar, Cartagena, Colombia.

## Abstract

We recently showed that dispersion-correcting potentials (DCPs), atom-centered Gaussian-type functions developed for use with B3LYP (*J. Phys. Chem. Lett.* **2012**, *3*, 1738–1744) greatly improved the ability of the underlying functional to predict non-covalent interactions. However, the application of B3LYP-DCP for the β-scission of the cumyloxyl radical led a calculated barrier height that was over-estimated by ca. 8 kcal/mol. We show in the present work that the source of this error arises from the previously developed carbon atom DCPs, which erroneously alters the electron density in the C-C covalent-bonding region. In this work, we present a new C-DCP with a form that was expected to influence the electron density farther from the nucleus. Tests of the new C-DCP, with previously published H-, N- and O-DCPs, with B3LYP-DCP/6-31+G(2d,2p) on the S66, S22B, HSG-A, and HC12 databases of non-covalently interacting dimers showed that it is one of the most accurate methods available for treating intermolecular interactions, giving mean absolute errors (MAEs) of 0.19, 0.27, 0.16, and 0.18 kcal/mol, respectively. Additional testing on the S12L database of complexation systems gave an MAE of 2.6 kcal/mol, showing that the B3LYP-DCP/6-31+G(2d,2p) approach is one of the best-performing and feasible methods for treating large systems dominated by non-covalent interactions. Finally, we showed that C-C making/breaking chemistry is well-predicted using the newly developed DCPs. In addition to predicting a barrier height for the β-scission of the cumyloxyl radical that is within 1.7 kcal/mol of the high-level value, application of B3LYP-DCP/6-31+G(2d,2p) to 10 databases that include reaction barrier heights and energies, isomerization energies and relative conformation energies gives performance that is amongst the best of all available dispersion-corrected density-functional theory approaches.

*Author of correspondence. Phone: +1-780-782-6672, E-mail: Gino.DiLabio@nrc.ca.

## Introduction

The ability to model structures and properties of systems in which non-covalent interactions are important is critical in a number of areas of chemistry, physics and biology, as is illustrated by the complexities of protein structure.[1,2] The central challenge associated with systems that are strongly influenced by van der Waals interactions is that these forces notoriously difficult to accurately reproduce using most electronic structure simulation techniques. One exception to this is complete basis set extrapolated coupled cluster with single and double excitations, with the non-iterative inclusion of corrections for triple excitations, i.e. CCSD(T)/CBS. This high-level correlated quantum mechanical wavefunction technique has been proven to be capable of describing non-covalent interactions with great accuracy. However, this method has limited applicability because it requires significant computational time and resources to treat even the smallest systems.

On the other hand, computational methods based on approximations to density-functional theory (DFT) have the potential to provide computationally efficient approaches to modeling systems in which non-covalent interactions play a central role. A critical deficiency of conventional DFT approximations is their well-known inability to accurately and reliably model non-covalent (especially dispersion) interactions. Several general approaches to improving the ability of DFT-based methods to describe non-covalent interactions have been developed. For example, early efforts focussed on the use of empirical, pair-wise dispersion parameters in conjunction with density functionals,[3,4] an approach that has been popularized largely thanks to the work of Grimme's group.[5,6,7] A very accurate approach to this was developed by Becke and Johnson, whose method calculates dispersion coefficients from the dipole moment of the exchange-hole.[8,9,10] Recently, Tkachenko and Scheffler developed a parameter-free approach to obtaining van der Waals parameters[11] that can be used to provide accurate non-covalent binding energies.[12] The common thread that associates these approaches is that they correct the DFT-predicted energy, not the underlying electronic structure predicted by the methods themselves. On the other hand, Zhao and Truhlar have developed several suites of "Minnesota" functionals that are capable of modeling a wide range of non-covalent interactions.[13,14,]

Our own solution to the DFT "dispersion problem" is atom-centered dispersion-correcting potentials (DCPs), which improve the long-range behaviour of the functional for which they were developed. [15,16,17,18,19] The DCP approach is similar in spirit to a plane wave-based technique introduced by von Lilienfeld and co-workers.[20,21,22] DCPs are atom-centered Gaussian-type functions, similar in structure to those used in effective-core potentials,[23] that correct the potential in which the electrons move. The main design principle behind DCP development is to adjust the erroneous long-range electron density predicted by a particular DFT-based method in non-covalently bonded systems such that non-covalent interactions are accurately reproduced. This is achieved by optimizing the exponents and coefficients of a set of Gaussian functions to reproduce the binding energies (BEs) associated with a fitting set of potential energy surfaces in of fitting sets composed of non-covalently interacting dimers.

We recently demonstrated that the DCP approach can be combined with the popular B3[24]LYP[25] functional and 6-31+G(2d,2p) basis sets to allow for the very accurate prediction structures and BEs for a wide variety of non-covalently interacting dimer systems.[19] Although the focus of that work was strictly on intermolecular binding, our broader interests lay in modeling chemical systems. Recent applications of the B3LYP-DCP approach guided the development of our understanding of how hydrogen bonding influences the mechanism and kinetics hydrogen atom transfer reactions between oxygen-centered radicals and various C, H, N, and O containing substrates by predicting rate constants that are in very good agreement with those measured by laser flash photolysis.[26,27,28] However, over the course of studying the C-C bond breaking process that occurs in the unimolecular β-scission reactions of the cumyloxyl ($C_6H_5C(CH_3)_2O^{\bullet}$) radical, we found that B3LYP-DCP/6-31+G(2d,2p) predicted barrier height of ca. 20.8 kcal/mol (vide infra), which is roughly 8 kcal/mol too high. This implied that the DCPs associated with the carbon atoms (C-DCPs) were over-stabilizing the radical as compared to the transition state structure. In other words, the C-DCPs were not simply influencing the potential at large distance from the nuclei (that is, at van der Waals distances), but also strongly influencing the potential on length scales commensurate with those of covalent bond lengths.

An examination of the form of the DCPs we reported in reference 19 revealed the source of the discrepancy arising from the comparison of the very good results we obtained for hydrogen atom transfer reactions in C, H, N, and O containing species along with the BEs for non-covalently bound dimers, against the poor results we observed for C-C β-scission: The C-DCPs contain some Gaussian functions that have coefficients that are an order of magnitude larger than those of the functions in the DCPs of other atoms. This led us to hypothesize that these larger coefficients result in significant influence of these Gaussian functions in the "covalent" region, thus effecting carbon-carbon bonds and thereby leading to the erroneously high β-scission barrier height. Since covalent properties were not incorporated into the fitting of DCPs, this short-coming of the C-DCPs initially escaped our attention.

In this work, we test the hypothesis that DCPs that are effective for describing the binding in non-covalently interacting dimer systems *and* C-C bond chemistry should be composed of Gaussian-type functions with coefficients that are a factor of 5-10 smaller than those associated with the C-DCPs published in reference 19 (that is, closer to the coefficients in associated with the DCPs produced for other atoms). Rather than incorporating C-C reaction data into the DCP optimization process to mitigate errors in C-C reaction chemistry, we generate new carbon DCPs by simply guiding their optimization toward smaller coefficients. It will be shown that the new C-DCPs so generated remain capable of predicting accurate binding energies in non-covalently interacting dimer systems while being well-behaved for carbon-carbon reaction chemistries and generally improve the performance of the B3LYP functional in the prediction of properties of a large set of covalently and non-covalently bound systems, as embodied in a number of commonly available benchmarks.

**Theoretical Background and Computational Approaches**

As described above, DCPs are atom-centered potentials (*U(r)*) composed of Gaussian-type functions that have the same form, and are utilized in the same way as, effective core potentials,[23] viz.:

$$U_l(r) = r^{-2} \sum_{i=1}^{N_l} c_{li} r^{n_{li}} e^{-\zeta_{li} r^2} \qquad (1)$$

In eq. (1), $l$ is the angular moment, $N_l$ is the number of Gaussian functions, $n_{li}$ is the power of $r$ (set to 2 throughout this work), $c_{li}$ is the coefficient of the Gaussian and $\xi_{li}$ is its exponent.

The goal of previous DCP optimization efforts was to improve the geometries and BEs of non-covalently bonded dimer and n-mer systems.[29] In the present work, we keep in mind this goal but develop carbon DCPs composed of Gaussian functions having coefficients that have magnitudes lower than ca. 0.005. In this way, we expect to be able to produce C-DCPs that can provide an accurate of non-covalent binding while not negatively impacting the properties of carbon-containing species in the covalent region.

The DCPs presented in this work were developed by optimizing $\xi_{li}$ and $c_{li}$ values (eq. 1) for carbon. We started with the C-DCP that was developed for use without corrections for basis set incompleteness by the counterpoise (CP) approach[30] as reported in reference 19. This initial DCP was modified so that the functions having coefficients with magnitudes larger than 0.01 were reduced by a factor of ten.[31] The optimizations were performed through the minimization of the error in the predicted BEs along one-dimensional potential energy surfaces for slipped-parallel and T-shaped benzene dimers, the ethane dimer, and the $D_{3h}$ and $C_{3v}$ conformations of the methane dimer. The use of fitting data that span a range of conformer geometries that include a path to dimer dissociation ensures that dissociation behavior (i.e. long-range behavior) is adequately reproduced. All fitting data are of CCSD(T)/CBS quality. The fitting data for the benzene dimers were kindly provided to us by Professor C. David Sherrill and are given in the Electronic Supplementary Material (ESM) Section. We also list the PES data for the ethane dimer in the ESM, and these data were obtained by varying the distance between rigid, B3LYP/6-31G(d)-optimized monomers and computing the reference binding energies by performing CCSD(T)/CBS calculations according to an approach we used elsewhere.[32,33] The data for the methane dimers were provided in the Supporting Information Section of reference 29. DCPs are also required for the hydrogen atom: In the present work, we used the H-DCPs given in reference 19 without alteration or further optimization.

In order to generate the DCPs, we performed a stochastic search in the parameter space defined by the $\xi_{li}$ and $c_{li}$ values associated with the carbon DCP functions so as to minimize the mean absolute error (MAE) of the calculated binding energies of in the fitting set according to Equation 2:

$$\mathrm{MAE} = \frac{1}{M}\sum_{1}^{M}\left(\frac{1}{N}\sum_{1}^{N}\left|\mathrm{BE}_i^{\mathrm{Ref}} - \mathrm{BE}_i^{\mathrm{DCP}}\right|\right) \quad (2)$$

Here, *N* accounts for the number points associated with each of the PESs in the fitting set, M is the number of PESs, and $BE^{\mathrm{Ref}}$ and $BE^{\mathrm{DCP}}$ represent the binding energies calculated using the high-level approach and DCPs, respectively. To avoid getting trapped in local minima associated with the DCP parameters, the error was allowed to increase a small amount for up to three consecutive steps, otherwise the optimization was restarted with slightly modified DCP parameters.

The DCP optimizations were performed using internally developed scripts that handled job submissions and data processing. Gaussian-09[34] was used to compute DFT energies in all cases. Because DCPs resemble effective core potentials, they can be used directly by the Gaussian package (and other computational chemistry programs) through a simple modification of input files. A utility program for generating Gaussian and NWChem input files with DCPs is available on our website[35] and a sample input file is provided in the ESM. We emphasize that the DCPs were optimized for use with B3LYP/6-31+G(2d,2p) without correction for basis set incompleteness. The same results will not be achieved using other basis sets or, for example, CP corrections to mitigate the effects of basis set incompleteness. Reference 19 provides more detail on the performance of the B3LYP-DCP approach with other basis sets. Likewise, the DCPs cannot be transferred to other functionals whilst maintaining the same degree of performance.

Throughout this work, the presently optimized C-DCPs were employed with previously developed H-, N-, and O-DCPs[19] to assess the performance of B3LYP-DCP/6-31+G(2d,2p) approach against several benchmarks of non-covalently bonded dimers. These benchmark sets include the S66[36] and S22B[37,38] sets of Hobza's group,[39] and the HSG-A set of Marshall et al.,[38] all of which provide representations of interactions that have relevance in biological systems. In addition, we tested the C-DCPs on the benchmark set of twelve dimers of saturated and unsaturated hydrocarbon molecules discussed by Granatier et al. (hereafter HC12).[40] We also applied the DCPs to the benchmark set of 12 supramolecular complexes (S12L) recently described by Grimme.[41]

To assess the performance of the DCPs on C-C reaction chemistry specifically, and intramolecular interactions in general, we applied them to subsets of systems containing the H, C, N, and O atoms within sub-databases of the very extensive GMTKN30 database recently presented by Goerigk and Grimme[42].[43]

The Gaussian-03[44] and Gaussian-09[34] packages were used for all of the calculations. Default integration grids were used throughout and the 6-31+G(2d,2p) basis sets employed pure (i.e. five primitive) d-type functions.

## Results and Discussion

### *Optimization of Carbon Dispersion-Correcting Potentials.*

The C-DCPs optimized for use with B3LYP/6-31+G(2d,2p) without corrections for basis set incompleteness are provided in Table 1. The optimization procedure was relatively straightforward. The errors calculating by Eq. 2 for the starting DCP was greater than 1.5 kcal/mol but decreased to acceptable values in short order. This suggests to us that the parameter space defined by the exponents and coefficients of the Gaussian functions comprising the DCPs have multiple minima corresponding to DCP parameters capable of good descriptions of non-covalent interactions.

The DCP in Table 1 have exponents that range between ca. 0.02 and 0.13 and coefficients with absolute values that are lower than 0.0032. For comparison, the optimized Gaussian functions of the C-DCPs reported in reference 19 have exponents ranging from ca. 0.01 to 0.20 and coefficients with absolute values over a range of 0.00002 – 0.057. We hypothesized that the lower coefficients associated with the DCPs presented in Table 1 are expected to lessen the negative impact of the DCPs on the covalent region of C-C bonds as compared to the C-DCPs of reference 19.

**Table 1.** Carbon dispersion-correcting potentials (DCPs) optimized for use with the B3LYP/6-31+G(2d,2p) density-functional theory-based method without counterpoise corrections for basis set incompleteness.[a]

| Function type | $\zeta_i$ | $c_i$ |
|---|---|---|
| F and higher | 0.091556053 | 0.000025303 |
| | 0.044472350 | 0.000137829 |
| | 0.019075560 | -0.000000056 |
| S-F | 0.075790561 | 0.000003145 |
| | 0.039119707 | 0.001009080 |
| P-F | 0.131194450 | -0.000000531 |
| | 0.045246336 | -0.003143976 |
| D-F | 0.033941983 | -0.002000967 |

[a]Additional formatting is required in order to use this DCPs in computational chemistry programs, see the ESM.

*Application of Dispersion-Correcting Potentials to the β-scission of Cumyloxyl*

To further explore how the C-DCP impact the covalent region of C-C bonds, we computed the barrier heights associated with the β-scission of the cumyloxyl radical,[45,46] the process for which is illustrated in Figure 1.

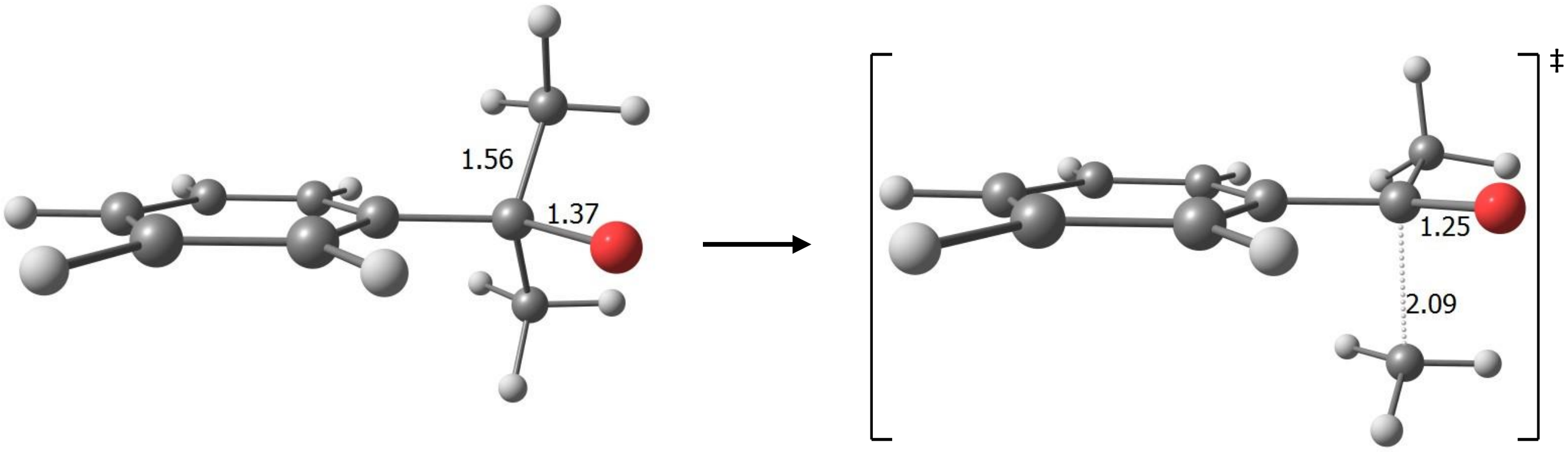

**Figure 1**. Calculated (B3LYP/6-31+G(2d,2p)) structures associated with the gas-phase β-scission of the cumyloxyl radical. Distances shown are in Å. Color key: carbon = gray, hydrogen = white, oxygen = red.

We calculated the geometries of the cumyloxyl radical and its associated β-scission transition state (TS) at the B3LYP/6-31+G(2d,2p) level of theory. The minimum energy structure had all positive frequencies while the TS structure had associated with it a single imaginary mode that represented the C-C bond breaking process that connected the cumyloxyl radical to benzaldehyde and methyl radical products. Figure 1 shows some key distances associated with the two structures. We then performed single-point calculations using the CBS-QB3 composite approach,[47] which is capable of predicting fairly accurate barrier heights. This procedure provided our benchmark, electronic energy barrier height of 13.1 kcal/mol. Using unadorned B3LYP/6-31+G(2d,2p) we obtain a barrier height of 11.7 kcal/mol, which is 1.4 kcal/mol too low. This value remains unchanged when pair-wise correction schemes for dispersion are employed because the bond changes involved in TS structure formation are below the cut-off distances usually employed in such schemes. On the other hand, the DCPs published in reference 19, as mentioned above, give a barrier height of 20.8 kcal/mol, which is 7.9 kcal/mol too high. Utilizing the carbon DCP given in Table 1, along with the H- and O-DCP of reference 19, gives a barrier height of 14.8 kcal/mol, which is too high by 1.7 kcal/mol. We reiterate that this latter result was achieved by simply guiding the C-DCP optimization toward smaller magnitude coefficients for Gaussian-type functions as described above and not by inclusion of the cumyloxyl radical and its β-scission transition state in the optimization process.

On the basis of the results for the β-scission barrier height, it is clear that our hypothesis that the C-DCPs from reference 19 over-stabilize the cumyloxyl radical relative to the β-scission transition state compared to both unadorned B3LYP and to CBS-QB3 is correct. The C-DCPs optimized in this work do appear to offer some preferential stabilization of the transition state, but to an extent that the predicted barrier height has an error that is on par with that given by B3LYP without DCPs.

*Application of Dispersion-Correcting Potentials to Intermolecular Non-covalently Interacting Systems*

In Tables 2-5, we illustrate the performance of the C-DCPs of the present work, along with the H, N, and O DCPs taken from reference 19, with respect to the treatment of non-covalently bound (nominally) dimer systems. Table 2 lists the high-level BEs of the S66 set along with the signed errors of the calculated BEs using the B3LYP-DCP/6-31+G(2d,2p) approach, and shows that the DCPs perform quite well. In general, the non-CP DCPs give a slight over-binding (mean signed error, MSE, = 0.09 kcal/mol) of the H-bonded dimers and a slight under-binding in dispersion-bound dimers (MSE = -0.05 kcal/mol). The dimers in the set that is characterized as interacting via mixed forces have an MSE = -0.03 kcal/mol.

**Table 2**. Signed errors in binding energies (SE) calculated for the non-covalently bonded dimers of the S66 benchmark set using B3LYP/6-31+G(2d,2p) with dispersion-correcting potentials without (Non-CP) counterpoise corrections. The data are ordered according to interaction type, with performance statistics provided for data in each section. Overall performance statistics for the entire S66 set are provided at the bottom of the Table. All data are in kcal/mol except percent values.

| Dimer | Accepted High-Level Value[a] | SE[b] |
|---|---|---|
| | *Hydrogen Bonded Dimers*[c] | |
| water – water | 5.03 | 0.32 |
| water – methanol | 5.71 | 0.14 |
| water – methylamine | 6.99 | 0.69 |
| water – peptide[d] | 8.21 | 0.01 |
| methanol – methanol | 5.85 | 0.07 |
| methanol – methylamine | 7.61 | 0.59 |
| methanol – peptide[d] | 8.14 | 0.15 |
| methanol – water | 5.10 | 0.25 |
| methylamine – methanol | 3.09 | 0.08 |
| methylamine – methylamine | 4.16 | 0.08 |
| methylamine – peptide[d] | 5.46 | -0.10 |
| methylamine – water | 7.37 | 0.50 |
| peptide[d] – methanol | 6.24 | -0.16 |
| peptide[d] – methylamine | 7.47 | 0.32 |
| peptide[d] – peptide[d] | 8.69 | -0.22 |
| peptide[d] – water | 5.20 | 0.06 |
| uracil – uracil (base pair) | 17.41 | -0.40 |
| water – pyridine | 6.97 | 0.46 |
| methanol – pyridine | 7.49 | 0.37 |
| acetic acid – acetic acid | 19.44 | -0.24 |
| acetamide – acetamide | 16.58 | -0.36 |
| acetic acid – uracil | 19.77 | -0.31 |
| acetamide – uracil | 19.40 | -0.20 |
| Mean Absolute/Signed Error | | 0.27/0.09 |
| Mean Absolute/Signed % Error | | 3.4/2.1 |
| *Dispersion-Dominated Dimers* | | |
| benzene – benzene[e] (π-stacked) | 2.82 | 0.19 |
| pyridine – pyridine (π-stacked) | 3.90 | 0.03 |
| uracil – uracil (π-stacked) | 9.83 | -0.32 |
| benzene – pyridine (π-stacked) | 3.44 | 0.07 |
| benzene – uracil (π-stacked) | 5.71 | -0.34 |
| pyridine – uracil (π-stacked) | 6.82 | -0.38 |
| benzene – ethane | 1.43 | 0.23 |
| uracil – ethane | 3.38 | 0.10 |
| uracil – ethyne | 3.74 | 0.16 |
| pyridine – ethane | 1.87 | 0.16 |
| pentane – pentane | 3.78 | -0.10 |
| neopentane – pentane | 2.61 | -0.13 |
| neopentane – neopentane | 1.78 | -0.04 |
| cyclopentane – neopentane | 2.40 | -0.05 |
| cyclopentane – cyclopentane | 3.00 | 0.00 |
| benzene – cyclopentane | 3.58 | 0.06 |
| benzene – neopentane | 2.90 | 0.09 |
| uracil – pentane | 4.85 | -0.25 |
| uracil – cyclopentane | 4.14 | -0.24 |
| uracil – neopentane | 3.71 | -0.04 |
| ethene – pentane | 2.01 | -0.12 |
| ethyne – pentane | 1.75 | 0.06 |
| peptide[c] – pentane | 4.26 | -0.20 |
| Mean Absolute/Signed Error | | 0.15/-0.05 |
| Mean Absolute/Signed % Error | | 4.3/0.0 |
| *Mixed Interactions* | | |
| benzene – benzene (T-shaped) | 2.88 | 0.08 |
| pyridine – pyridine (T-shaped) | 3.54 | -0.15 |
| benzene – pyridine (T-shaped) | 3.33 | 0.03 |
| benzene – ethyne (CH-π) | 2.87 | 0.19 |
| ethyne – ethyne (T-shaped) | 1.52 | 0.30 |
| benzene – acetic acid (OH-π) | 4.71 | -0.30 |
| benzene – formamide (NH-π) | 4.36 | -0.13 |
| benzene – water (OH-π) | 3.28 | 0.16 |
| benzene – methanol (OH-π) | 4.19 | 0.05 |
| benzene – methylamine (OH-π) | 3.23 | -0.04 |
| benzene – peptide (NH-π) | 5.28 | -0.16 |
| pyridine – pyridine (CH-N)[f] | 4.15 | -0.36 |
| ethyne – water (CH-O)[f] | 2.85 | 0.29 |
| ethyne – acetic acid (OH- π)[f] | 4.87 | 0.34 |
| pentane – acetic acid | 2.91 | -0.03 |
| pentane – acetamide | 3.53 | -0.18 |
| benzene – acetic acid (π-stacked) | 3.80 | 0.09 |
| peptide – ethane | 3.00 | 0.00 |
| methylamine – pyridine[g] | 3.97 | -0.04 |
| pyridine – ethyne | 3.99 | 0.40 |
| Mean Absolute/Signed Error | | 0.17/0.03 |
| Mean Absolute/Signed % Error | | 5.0/1.6 |
| *Performance Statistics for the Entire S66 Set* | | |
| Mean Absolute Error | | 0.19 |
| Mean Signed Error | | 0.01 |
| Mean Absolute Percent Error | | 4.17 |
| Mean Signed Percent Error | | 0.99 |
| Error Range | | -0.40 to +0.69 |
| Percent Error Range | | -8.8 to +19.7 |

[a]Reference values are generally of CCSD(T)/CBS quality. Revised values associated with the set of hydrogen-bonded dimers were taken from reference [48] and are taken from reference **Error! Bookmark not defined.**. [b]Defined as BE(DCP) – BE(Reference). [c]All indicated dimers have a hydrogen-bonded interaction in the form *donor-acceptor*, with the exception of methylamine-water and dimers in which more than one hydrogen bond are formed.[d]"Peptide" is the nomenclature used in reference **Error! Bookmark not defined.** and refers to methylamide. [e]Slipped-parallel configuration. [f]Hydrogen-bonding interaction. [g]NH-N and CH(methylamine)-π interactions.

Overall, the performance of the B3LYP-DCP/6-31+G(2d,2p) is excellent, with an MAE of 0.19 kcal/mol. This MAE can be compared to the value of 0.14 kcal/mol for the DCPs of reference 19. Approximately half of the 0.05 kcal/mol difference in MAE for the S66 set arises from the use of updated reference BEs for the hydrogen-bonded set of dimers, see reference 48. Additional comparisons can be made to other dispersion-corrected B3LYP approaches. For example, the B3LYP-D3/def2-QZVP approach[49] developed by Grimme et al. gives an MAE of 0.40 kcal/mol when the D3 approach is used in conjunction with their damping function,[50] and this MAE drops to 0.28 kcal/mol when the damping function of Becke and Johnson[8,9,10] is utilized. We note that better performance for the S66 set is achieved using other functionals in conjunction with D3 corrections, see reference 50.

The DCPs perform in a similar fashion on the S22B set of dimers, which, to some extent, is a subset of the dimers in the S66 set. As such, we relegate the detailed data to the ESM and report herein overall performance statistics. The present DCPs give an MAE for the S22B set of 0.27 kcal/mol, an increase of 0.04 kcal/mol over the performance of the DCPs in reference 19. Despite the increase in MAE, the B3LYP-DCP approach nevertheless performs better than B3LYP-D3/def2-QZVP for this set, for which an MAE of 0.36 kcal/mol is obtained.[43] For the interested reader, reference 19 contains comparisons to other dispersion-corrected DFT-based methods.

Additional performance assessments of B3LYP-DCP/6-31+G(2d,2p) was made by applying the method to the HSG-A benchmark set. This set of non-covalently bound dimers and trimers is composed of 21 neutral and charged fragments obtained from the HIV-II protease crystal structure having a bound inhibitor molecule, indinavir. To some extent, this set is a more stringent test of dispersion-corrected DFT-based methods because the presence of charges introduces polarization effects. High-level binding energies were reported by Faver et al.[51] and recently refined by Marshall et al.[52] For our calculations, we used structures graciously provided to us by Professor David Sherrill.

As can be seen in Table 3, the B3LYP-DCP/6-31+G(2d,2p) approach performs well for the HSG-A set and, in fact, shows a small improvement over the performance of the DCPs in reference 19 . The largest error in the BEs is predicted for the acetate ion – water – acetic acid trimer is 1.16 kcal/mol, which represents an error of 6.2%. Despite this error, we obtain an MAE for the HSG-A set of 0.16 kcal/mol. Burns et al.[52] performed an extensive review of dispersion-enabled DFT-based methods used with aug-ccpVTZ basis sets and they reported the following MAEs, in kcal/mol: B3LYP-XDM = 0.14, B3LYP-D3 = 0.48, M06-2X = 0.47 and ωB97X-D = 0.32.

**Table 3**: Signed errors in binding energies (SE) calculated for the non-covalently bonded systems of the HSG-A set using B3LYP/6-31+G(2d,2p) with dispersion-correcting potentials without (Non-CP) counterpoise corrections. Statistics associated with the performance of the methods are also provided. All values except percent are in kcal/mol.

| Complex | HSG-A Name | Accepted High-Level Value[a] | SE[b] |
|---|---|---|---|
| methane – *N*-*t*-butylformamide | ala29-big | 0.518 | -0.03 |
| ethane – indan-2-ol | ala128-small | 2.283 | 0.22 |
| benzene – *N*-methylguanidine($H^+$)[c] | arg8 | 2.478 | -0.02 |
| acetate ion – water – acetic acid | ash26-asp125 | 16.526 | -1.16 |
| 3-acetylamino-propioniate ion – methanol | asp129-big | 19.076 | -0.29 |
| benzene – acetate ion | asp130 | 5.998 | -0.04 |
| dimethylaminomethanol – 2-FAA[d] | gly28-big | 3.308 | -0.05 |
| Pyridine – 2-formylaminoacetamide | gly50-ring-big | 0.581 | -0.26 |
| *N*-methylacetamide – 2-FAA | gly50-v1 | 5.066 | 0.07 |
| *N*-methylacetamide – *N*-methylacetamide | gly127 | 7.509 | -0.25 |
| *N*-methylacetamide – 2-FAA | gly148 | 6.274 | -0.05 |
| propane – *N*-*t*-butylformamide | ile48-big | -0.302 | 0.11 |
| Butane – benzene | ile147 | 2.103 | 0.09 |
| butane – *N*-*t*-butylformamide | ile150-big | 1.378 | -0.04 |
| ethane – ethane-1,2-diamine | ile184 | 0.856 | -0.07 |
| toluene – 2-methylpropane | leu23-big | 1.100 | -0.04 |
| ethane – 3-methylpyridine | pro181 | 1.534 | 0.06 |
| ethane – *N*-*t*-butylformamide | val33-big | 0.472 | -0.01 |
| propane – benzene | val83 | 1.598 | 0.25 |
| propane – benzene | val132 | -0.378 | -0.10 |
| acetate ion – water – *N*-methylacetamide | wat200 | 9.538 | -0.09 |
| Mean Absolute/Signed Error | | | 0.16/-0.08 |
| Mean Absolute/Signed % Error | | | 3.7/-6.0 |

[a]Taken from reference 52. [b]Defined as BE(DCP) – BE(Reference). [c]Protonated *N*-methylguanidine. [d]2-formylaminoacetamide.

Next, we consider the "HC12" set recently described by Granatier.[40] This set contains dimers of saturated linear and cyclic species and unsaturated linear and cycle hydrocarbons. The B3LYP-DCP/6-31+G(2d,2p) results presented in Table 4 demonstrates fairly good performance for this set of dimers, giving an MAE of 0.18 kcal/mol, which corresponds to a mean absolute percent error of 6.1. Granatier et al. reported BEs for a number of dispersion corrected DFT-based methods but these did not include the B3LYP functional. The best "D3" method reported in reference 40 was TPSS-D3/def2-QZVP, which gave a very low MAE of 0.14 kcal/mol, while BLYP-D3/def2-QZVP produced an MAE of 0.56 kcal/mol. M06-2X/aug-cc-pVTZ displays very good performance for the HC12 set, as evidenced by an MAE of 0.15 kcal/mol.

**Table 4**: Signed errorss in binding energies (SD) calculated for the non-covalently bonded hydrocarbon dimers of the HC12 set using B3LYP/6-31+G(2d,2p) with dispersion-correcting potentials without (Non-CP) counterpoise corrections. Statistics associated with the performance of the methods are also provided. All values except percent are in kcal/mol.

| Dimer | Accepted High-Level Value[a] | SE[b] |
|---|---|---|
| | *Saturated linear hydrocarbons*[c] | |
| propane – propane | 2.055 | -0.11 |
| butane – butane | 2.912 | -0.10 |
| pentane – pentane | 3.772 | -0.17 |
| hexane – hexane | 4.649 | -0.17 |
| *Saturated cyclic hydrocarbons* | | |
| cyclopropane – cyclopropane | 2.219 | -0.20 |
| cyclobutane – cyclobutane | 2.943 | -0.15 |
| cyclopentane – cyclopentane | 3.433 | -0.25 |
| cyclohexane – cyclohexane | 2.743 | -0.03 |
| *Unsaturated linear and cyclic hydrocarbons* | | |
| butadiene – butadiene | 2.273 | -0.10 |
| hexatriene – hexatriene | 4.497 | -0.32 |
| cyclobutadiene – cyclobutadiene | 1.995 | 0.33 |
| benzene – benzene | 2.732 | 0.19 |
| Mean absolute/signed error | | 0.18/-0.09 |
| Mean absolute/signed % error | | 6.1/-2.2 |

[a]The reference BEs are non-CP corrected, CCSD(T)/CBS values taken from reference 53. [b]Defined as BE(DCP) – BE(Reference).

Finally, we consider the S12L database of Grimme.[41] Grimme performed careful calculations on this set of experimentally studied systems in order to assess the contributions to the measured free-energies of binding of solvent and molecular vibrations to ultimately arrive at BEs. The set includes two "tweezer" complexes with tetracyanoquinone (TCNQ) and 1,4-dicyanobenzene ( complexes 2a and 2b), two "pincer" species complexed with heteroatom-substitute π-delocalized molecules, (complexes 3a and 3b), a "buckycatcher" species complexed with $C_{60}$ and $C_{70}$ fullerenes (complexes 4a and 4b), an amide macrocycle coupled with benzoquinone and glycine glycine anhydride (complexes 5a and 5b), complexes of cucurbit[6]uril cation with butylammonium and propylammonium (complexes 6a and 6b), and finally complexes of cucurbit[7]uril bis(trimethylammoniomethyl) ferrocene and with neutral 1-hydroxyadamantane (complexes 7a and 7b). The reader is directed to reference 41 for additional details of the complexes, including diagrams of the structures. There are two cases where the complexes contain atoms for which we do not yet have DCPs, that is, complex 3b (chlorine) and complex 7a (iron). For these two cases, we simply apply DCPs to all of the C, H, N and O atoms and leave the Cl and Fe atoms unadorned. Our expectations are that the omission of DCPs for these centers will have little or no impact on the overall binding energies we calculate, in particular for the complex 7a, where the iron is buried inside of the ferrocenyl moiety.

The results for the S12L benchmark set are presented in Table 5. In general, the errors between the B3LYP-DCP calculated values and the accepted BEs are larger than those associated with other benchmark sets representing non-covalent interactions in dimer systems. The larger discrepancies can be expected on the basis that approximate methods, in particular a continuum solvent model, were used to back-correct experimentally measured binding free-energies to obtain gas-phase BEs. Nevertheless, the MAE of 2.6 kcal/mol produced by our DCP approach is very good, in particular when compared to other dispersion-corrected DFT-based methods.[55] Although B3LYP-based methods were not explored in reference 55, they did report on a number of other dispersion-corrected methods: The best performing approach was one dubbed PBE-D2/QZVP'+$E^{ABC}$. This approach, which gives an MAE of 1.5 kcal/mol, uses older generation pair-wise dispersion coefficients along with the PBE[54] functional with quadruple-zeta basis sets and corrections for three-body terms. Of the "workable" approaches examined in reference 55, PBE-D2 achieved an MAE of 1.6 kcal/mol when used with triple-zeta basis sets and CP corrections, while the best performing D3-based approach was found to be PBE-D3/TZVP with CP corrections, which gave an MAE of 2.3 kcal/mol. For comparison, the M06-2X/TZVP approach with CP corrections gave an MAE of 3.3 kcal/mol.

**Table 5**: Signed errors in binding energies (SE) calculated for the non-covalently bonded dimers of the S12L set using B3LYP/6-31+G(2d,2p) with dispersion-correcting potentials without counterpoise corrections. Statistics associated with the performance of the methods are also provided. All values except percent are in kcal/mol.

| Complex[a] | Accepted High-Level Value[b] | SE[c] |
|---|---|---|
| 2a | 29.9 | -0.5 |
| 2b | 20.5 | -1.1 |
| 3a | 24.3 | -1.9 |
| 3b | 20.4 | -2.7 |
| 4a | 27.5 | 3.1 |
| 4b | 28.7 | 3.5 |
| 5a | 34.8 | 0.0 |
| 5b | 21.3 | 0.9 |
| 6a | 77.4 | 8.1 |
| 6b | 77.0 | 4.7 |
| 7a | 131.5 | 0.9 |
| 7b | 22.6 | 6.1 |
| Mean absolute/signed error | | 2.6/1.6 |
| Mean absolute/signed % error | | 7.9/3.7 |

[a]See text and reference 41 for additional details related to the nature of the complexes. [b]Taken from reference 55. [c]Defined as BE(DCP) – BE(Reference).

We summarize this section by reiterating that the carbon atom DCP developed in this work, when used in conjunction with H-, N-, and O-DCPs provided in reference 19, and B3LYP/6-31+G(2d,2p) is a computationally efficient and accurate method for calculating the binding energies of non-covalently interacting dimer systems. We demonstrated that the performance of this approach is amongst the best of available dispersion-corrected DFT-based methods.

*Application of Dispersion-Correcting Potentials to a subset of a General Database of Thermochemistry, Kinetics and Non-covalent Interactions.*

As a final assessment of the B3LYP-DCP/6-31+G(2d,2p) approach, we apply it to a subset of the extensive "GMTKN30" database recently compiled by Goerigk and Grimme.[43] For brevity, we describe the database as consisting of sub-databases taken from various literature sources that contain high-level calculated or experimental data for a variety of thermochemical properties, reaction energies and barrier heights, and intra- and intermolecular non-covalent interactions. The interested reader is referred to reference 43 for a brief description of each sub-database. The GMTKN30 database is meant to provide a means for evaluating general methodologies, rather than exclusively dispersion-corrected DFT-based methods. It therefore contains databases that allow for the testing of the ability of methods on a number of different performance metrics. For the present work, we are particularly interested in inter- and intramolecular interactions as they impact non-covalent interactions and reaction chemistry.

We therefore selected a subset of 10 pertinent databases from the GMTKN30 database to test the B3LYP-DCP approach for their ability to predict properties based on these interactions. We performed calculations using the following sub-databses: BHPERI (barrier heights for pericyclic reactions),[56] DARC (Diels-Alder reaction energies),[57] BSR36 (bond separation energies of saturated hydrocarbons),[58] ISO34 (isomerization energies of small and medium sized organic molecules),[59] ISOL22 (isomerization energies of large organic molecules),[60] PCONF (relative energies of phenylalanyl-glycyl-gycine tripeptide conformers),[61] ACONF (relative energies of alkane conformers),[62] SCONF (relative energies of sugar conformers),[42,63] IDISP (intramolecular dispersion interactions),[42,43] and ADIM6 (interactions of n-alkane dimers).[49] A summary of the performance of B3LYP/6-31+G(2d,2p) without DCPs, with the DCPs of reference 19, and those developed in this work are given in Table 6 in the form of mean absolute errors relative to the reference data in each of the sub-databases. For comparison, we also present the results of calculations using B3LYP-D3/6-31+G(2d,2p). The explicit data for each of the relevant entries are provided in the ESM. The data presented in Table 6 and the ESM do not include data for species that contain elements other than H, C, N, and O atoms.

**Table 6**. Mean absolute errors (MAE, kcal/mol) of the B3LYP/6-31+G(2d,2p) approach on the sub-databases of the GMTKN30 database, without and with various dispersion-correction schemes. The lowest MAE for each database is indicated in bold.

| Sub-database name | B3LYP | B3LYP-DCP (ref. 19) | B3LYP-DCP (this work) | B3LYP-D3 |
|---|---|---|---|---|
| BHPERI | 3.44 | 3.36 | 2.48 | **1.67** |
| DARC | 10.71 | 13.02 | 4.44 | **3.54** |
| BSR36 | 9.77 | 3.29 | **1.34** | 2.41 |
| ISO34 | 2.02 | 3.16 | 1.60 | **1.49** |
| ISOL22 | 7.34 | 5.46 | **2.39** | 4.58 |
| PCONF | 3.71 | 0.64 | **0.46** | 0.57 |
| ACONF | 1.02 | 0.17 | **0.06** | 0.07 |
| SCONF | 0.93 | 1.07 | 0.77 | **0.36** |
| IDISP | 15.64 | 2.99 | **1.45** | 2.53 |
| ADIM6 | 4.81 | 0.14 | **0.09** | 0.34 |

The data in Table 6 shows that unadorned B3LYP/6-31+G(2d,2p) predicts fairly large MAEs for the 10 sub-databases. Particularly large MAEs are found for the ISOL22 and IDSIP databases. This is not surprising in light of the fact that the reference systems in these sub-databases are largely dominated by dispersion interactions, which B3LYP cannot model properly.  The Diels-Alder sub-database barrier heights, DARC, is also poorly treated by unadorned B3LYP, suggesting that short-range interactions that occur at the transition states are strongly influenced by non-covalent interactions.

The inclusion of dispersion corrections via the DCPs of reference 19 offers improvement to most, but not all, of the sub-databases. The MAEs for ISO34, DARC and SCONF *increase* with the use of the C-DCPs of reference 19 (as does that for SCONF), while that for BHPERI is only very slightly reduced. Large reductions in MAEs are seen for the remaining databases. These observations broaden the scope of systems that are poorly modeled with the C-DCPs of reference 19 and support our conclusion that they adversely affect the predicted energies associated with C-C bond making/breaking chemistry.

On the other hand, the C-DCPs developed in this work, used in conjunction with the H-, N-, and O-DCPs presented in reference 19, uniformly and significantly improve the MAEs of all 10 sub-databases. In particular, the databases associated with reaction barriers/energies - BHPERI, DARC and BSR36 – have their associated MAEs reduced by ca. 28, 59 and 86%, respectively.  The databases representing isomerization and relative conformation energies are have their MAES reduced by varying amounts, while those associated with intra- and intermolecular dispersion interactions (IDISP and ADMI6) are greatly improved.

The performance of the B3LYP-DCP approach of this work and that of the B3LYP-D3/6-31+G(2d,2p)  can be compared, with the proviso that the D3 approach was developed for use with def2-QZVP basis sets. Table 6 shows that B3LYP-D3 performs better, on average, for the BHPERI, DARC, PCONF and SCONF sub-databases, while the B3LYP-DCP performs better on the remaining six sub-databases. Although the B3LYP-DCP approach was parameterized using a very small number of non-covalently interacting dimer systems,[64] its excellent performance suggests that it is capturing a very broad range of intra- and intermolecular interactions.

Additionally, whereas the D3 approach fitted parameters using the S22, PCONF, SCONF, ACONF, CCONF, ADIM6 and other databases, it seems reasonable that general DCP performance can be improved further by broadening the fitting set to include constituents of the GMTKN30 database.

**Summary and Conclusions**

Dispersion-correcting potentials (DCPs) have been shown to be a very good approach for correcting the long-range behaviour of density-functionals for which they are designed, thereby greatly improving the ability of DFT-based methods to describe non-covalent intermolecular interactions. We recently developed a new set of DCPs for use with B3LYP that together provide very accurate predictions of geometries and binding energies (BEs) of non-covalently bonded dimers.[19] However, the recent application of our B3LYP-DCP approach to study the β-scission of the cumyloxyl radical led to a calculated barrier height that was over-estimated by ca. 8 kcal/mol. We hypothesized that the form of the previously developed carbon atom DCPs described in reference 19 was such that it has the unintended effect of erroneously altering the electron density in the covalent-bonding region, resulting in the over-estimation of the barrier for the C-C bond-breaking process.

In the present work, we developed a new C-DCP with a form that wass expected to more strongly influence the electron density farther from the nucleus and less so in the covalent region. Tests of the new C-DCP, in conjunction with previously published H-, N- and O-DCPs,[19] with B3LYP-DCP/6-31+G(2d,2p) on the S66, S22B, HSG-A, and HC12 databases of non-covalently interacting dimers showed that it is one of the most accurate methods available for treating intermolecular interactions: The B3LYP-DCP approach gave mean absolute errors (MAEs) of 0.19, 0.27, 0.16, and 0.18 kcal/mol, respectively, for these four databases. Additional testing on the S12L database of very large complexation systems gave an MAE of 2.6 kcal/mol, demonstrating that the B3LYP-DCP/6-31+G(2d,2p) approach is one of the best-performing and most feasible methods for treating large systems dominated by non-covalent interactions.

Finally, we showed that the modeling of C-C bond making/breaking chemistry is well-predicted using the newly developed DCPs. We applied B3LYP-DCP/6-31+G(2d,2p) to 10 databases that include reaction barrier heights and energies, isomerization energies, and relative conformation energies. We found our approach to perform very well on these systems, giving results that amongst the best of all available dispersion-corrected density-functional theory approaches. Furthermore, the new C-DCPs provide a more reasonable prediction of the barrier height for β-scission of the cumyloxyl radical.

Our findings demonstrate a number of important aspects of DPCs: 1) The parameter space defined by the Gaussian-type function exponents and coefficients of the DCPs has multiple minima capable of allowing for a good description of intermolecular interactions. 2) DCPs composed of functions with coefficients of small magnitude are capable of simultaneously providing a good description of inter- and intramolecular interactions. 3) Expansion of the relatively small fitting set used to derive DCPs to include other properties, such as reaction energies, is likely to be an effective means of further improving the performance of a density-functional in terms of accurate prediction of molecular properties, including reaction energetics and inter/intramolecular interactions. Our future efforts will focus on these aspects of DCP development.

**Acknowledgements**

The authors are grateful to the Centre for Oil Sands Innovation and WestGrid for providing generous support for this work. We thank Dr. Lars Goerigk for helpful discussions and Professor C. David Sherrill for providing some of the fitting data used in this work.

**Electronic Supplementary Material Available**

Potential energy surface structures and binding energy data used to fit the C-DCPs, a sample input file demonstrating the use of DCPs, the full table of data listing the calculated BEs associated with the S22 (“B”) dataset, the full data for the results listed in Table 6.

Electronic Supplementary Material

for

# Extension of the B3LYP – Dispersion-Correcting Potential Approach to the Accurate Treatment of both Inter- and Intramolecular Interactions.

Gino A. DiLabio[1,2 *] Mohammad Koleini,[1,3] and Edmanuel Torres[1,4]

[1]National Institute for Nanotechnology, National Research Council of Canada, 11421 Saskatchewan Drive, Edmonton, Alberta, Canada T6G 2M9; [2]Department of Physics, University of Alberta, Edmonton Alberta, Canada T6G 2E1; [3]Department of Chemical and Materials Engineering, University of Alberta, Edmonton Alberta, Canada T6G 2V4; [4]Faculty of Basic Sciences, Universidad Tecnológica de Bolívar, Cartagena, Colombia.

[*] Author of correspondence. Phone: +1-780-782-6672, E-mail: Gino.DiLabio@nrc.ca.

**Electronic Supplementary Material Available**

Potential energy surface structures and binding energy data used to fit the C-DCPs, a sample input file demonstrating the use of DCPs, the full table of data listing the calculated BEs associated with the S22 ("B") dataset, the full data for the results listed in Table 6.

Structures and binding energies of slipped-parallel (SP) and T-shaped (T) benzene dimer (data provided courtesy of Professor C. David Sherrill).

SP 1: BE = 0.15 kcal/mol

| | | | |
|---|---|---|---|
| C | 1.7459456310000001 | 0.9656274409999999 | 2.1437979999999999 |
| C | 1.4309664150000001 | 1.5875512599999997 | 3.3500030000000001 |
| C | 1.4309664150000001 | 1.5875512599999997 | 0.9375929999999999 |
| C | 0.8020805770000001 | 2.8313851569999997 | 3.3507520000000000 |
| C | 0.4893251290000001 | 3.4546152269999997 | 2.1437979999999999 |
| C | 0.8020805770000001 | 2.8313851569999997 | 0.9368439999999998 |
| H | 1.6808247000000001 | 1.1069077629999999 | 0.0014829999999999 |
| H | 0.5575053050000001 | 3.3137716629999998 | 0.0000000000000000 |
| H | 0.0000000000000000 | 4.4193718339999997 | 2.1437979999999999 |
| H | 0.5575053050000001 | 3.3137716629999998 | 4.2875959999999997 |
| H | 1.6808247000000001 | 1.1069077629999999 | 4.2861130000000003 |
| H | 2.2466158350000001 | 0.0072232569999997 | 2.1437979999999999 |
| C | 3.8765433480000002 | 3.4537443930000000 | 2.1437979999999999 |
| C | 4.1915225640000005 | 2.8318205740000000 | 0.9375929999999999 |
| C | 4.1915225640000005 | 2.8318205740000000 | 3.3500030000000001 |
| C | 4.8204084020000000 | 1.5879866770000000 | 0.9368439999999998 |
| C | 5.1331638509999999 | 0.9647566069999998 | 2.1437979999999999 |
| C | 4.8204084020000000 | 1.5879866770000000 | 3.3507520000000000 |
| H | 3.3758731439999998 | 4.4121485770000000 | 2.1437979999999999 |
| H | 3.9416642800000004 | 3.3124640710000000 | 4.2861130000000003 |
| H | 5.0649836740000005 | 1.1056001709999999 | 4.2875959999999997 |
| H | 5.6224889789999999 | 0.0000000000000000 | 2.1437979999999999 |
| H | 5.0649836740000005 | 1.1056001709999999 | 0.0000000000000000 |
| H | 3.9416642800000004 | 3.3124640710000000 | 0.0014829999999999 |

SP 2: BE = 2.81 kcal/mol

| | | | |
|---|---|---|---|
| C | 2.2552170000000000 | 0.8702309999999998 | 2.1437979999999999 |
| C | 1.8485389999999999 | 1.4364589999999997 | 3.3500030000000001 |
| C | 1.8485389999999999 | 1.4364589999999997 | 0.9375929999999999 |
| C | 1.0362450000000001 | 2.5690659999999998 | 3.3507520000000000 |
| C | 0.6315640000000000 | 3.1369259999999999 | 2.1437979999999999 |
| C | 1.0362450000000001 | 2.5690659999999998 | 0.9368439999999998 |
| H | 2.1691890000000003 | 0.9998459999999998 | 0.0014829999999999 |
| H | 0.7205480000000000 | 3.0082119999999999 | 0.0000000000000000 |
| H | 0.0000000000000000 | 4.0151750000000002 | 2.1437979999999999 |
| H | 0.7205480000000000 | 3.0082119999999999 | 4.2875959999999997 |
| H | 2.1691890000000003 | 0.9998459999999998 | 4.2861130000000003 |
| H | 2.8970170000000000 | 0.0000000000000000 | 2.1437979999999999 |
| C | 4.3508670000000000 | 3.7135790000000002 | 2.1437979999999999 |
| C | 4.7575450000000004 | 3.1473510000000000 | 0.9375929999999999 |
| C | 4.7575450000000004 | 3.1473510000000000 | 3.3500030000000001 |
| C | 5.5698390000000000 | 2.0147439999999999 | 0.9368439999999998 |
| C | 5.9745200000000001 | 1.4468839999999998 | 2.1437979999999999 |
| C | 5.5698390000000000 | 2.0147439999999999 | 3.3507520000000000 |
| H | 3.7090670000000001 | 4.5838099999999997 | 2.1437979999999999 |
| H | 4.4368949999999998 | 3.5839639999999999 | 4.2861130000000003 |
| H | 5.8855360000000001 | 1.5755979999999998 | 4.2875959999999997 |
| H | 6.6060840000000001 | 0.5686349999999998 | 2.1437979999999999 |
| H | 5.8855360000000001 | 1.5755979999999998 | 0.0000000000000000 |
| H | 4.4368949999999998 | 3.5839639999999999 | 0.0014829999999999 |

SP 3: BE = 1.92 kcal/mol

| | | | |
|---|---|---|---|
| C | 1.7459456310000001 | 0.9656274409999999 | 2.1437979999999999 |
| C | 1.4309664150000001 | 1.5875512599999997 | 3.3500030000000001 |
| C | 1.4309664150000001 | 1.5875512599999997 | 0.9375929999999999 |
| C | 0.8020805770000001 | 2.8313851569999997 | 3.3507520000000000 |
| C | 0.4893251290000001 | 3.4546152269999997 | 2.1437979999999999 |
| C | 0.8020805770000001 | 2.8313851569999997 | 0.9368439999999998 |
| H | 1.6808247000000001 | 1.1069077629999999 | 0.0014829999999999 |
| H | 0.5575053050000001 | 3.3137716629999998 | 0.0000000000000000 |
| H | 0.0000000000000000 | 4.4193718339999997 | 2.1437979999999999 |
| H | 0.5575053050000001 | 3.3137716629999998 | 4.2875959999999997 |
| H | 1.6808247000000001 | 1.1069077629999999 | 4.2861130000000003 |
| H | 2.2466158350000001 | 0.0072232569999997 | 2.1437979999999999 |
| C | 5.0061102589999997 | 3.4537443930000000 | 2.1437979999999999 |
| C | 5.3210894750000000 | 2.8318205740000000 | 0.9375929999999999 |
| C | 5.3210894750000000 | 2.8318205740000000 | 3.3500030000000001 |
| C | 5.9499753129999995 | 1.5879866770000000 | 0.9368439999999998 |
| C | 6.2627307619999995 | 0.9647566069999998 | 2.1437979999999999 |
| C | 5.9499753129999995 | 1.5879866770000000 | 3.3507520000000000 |
| H | 4.5054400550000002 | 4.4121485770000000 | 2.1437979999999999 |
| H | 5.0712311909999999 | 3.3124640710000000 | 4.2861130000000003 |
| H | 6.1945505850000000 | 1.1056001709999999 | 4.2875959999999997 |
| H | 6.7520558900000003 | 0.0000000000000000 | 2.1437979999999999 |
| H | 6.1945505850000000 | 1.1056001709999999 | 0.0000000000000000 |
| H | 5.0712311909999999 | 3.3124640710000000 | 0.0014829999999999 |

SP 4: BE = 0.53 kcal/mol

| | | | |
|---|---|---|---|
| C | 1.7459456310000001 | 0.9656274409999999 | 2.1437979999999999 |
| C | 1.4309664150000001 | 1.5875512599999997 | 3.3500030000000001 |
| C | 1.4309664150000001 | 1.5875512599999997 | 0.9375929999999999 |
| C | 0.8020805770000001 | 2.8313851569999997 | 3.3507520000000000 |
| C | 0.4893251290000001 | 3.4546152269999997 | 2.1437979999999999 |
| C | 0.8020805770000001 | 2.8313851569999997 | 0.9368439999999998 |
| H | 1.6808247000000001 | 1.1069077629999999 | 0.0014829999999999 |
| H | 0.5575053050000001 | 3.3137716629999998 | 0.0000000000000000 |
| H | 0.0000000000000000 | 4.4193718339999997 | 2.1437979999999999 |
| H | 0.5575053050000001 | 3.3137716629999998 | 4.2875959999999997 |
| H | 1.6808247000000001 | 1.1069077629999999 | 4.2861130000000003 |
| H | 2.2466158350000001 | 0.0072232569999997 | 2.1437979999999999 |
| C | 6.1356771700000001 | 3.4537443930000000 | 2.1437979999999999 |
| C | 6.4506563859999995 | 2.8318205740000000 | 0.9375929999999999 |
| C | 6.4506563859999995 | 2.8318205740000000 | 3.3500030000000001 |
| C | 7.0795422239999999 | 1.5879866770000000 | 0.9368439999999998 |
| C | 7.3922976729999998 | 0.9647566069999998 | 2.1437979999999999 |
| C | 7.0795422239999999 | 1.5879866770000000 | 3.3507520000000000 |
| H | 5.6350069659999997 | 4.4121485770000000 | 2.1437979999999999 |
| H | 6.2007981020000003 | 3.3124640710000000 | 4.2861130000000003 |
| H | 7.3241174959999995 | 1.1056001709999999 | 4.2875959999999997 |
| H | 7.8816228009999998 | 0.0000000000000000 | 2.1437979999999999 |
| H | 7.3241174959999995 | 1.1056001709999999 | 0.0000000000000000 |
| H | 6.2007981020000003 | 3.3124640710000000 | 0.0014829999999999 |

SP 5: BE = 0.07 kcal/mol

| | | | |
|---|---|---|---|
| C | 1.7459456310000001 | 0.9656274409999999 | 2.1437979999999999 |
| C | 1.4309664150000001 | 1.5875512599999997 | 3.3500030000000001 |
| C | 1.4309664150000001 | 1.5875512599999997 | 0.9375929999999999 |
| C | 0.8020805770000001 | 2.8313851569999997 | 3.3507520000000000 |
| C | 0.4893251290000001 | 3.4546152269999997 | 2.1437979999999999 |
| C | 0.8020805770000001 | 2.8313851569999997 | 0.9368439999999998 |
| H | 1.6808247000000001 | 1.1069077629999999 | 0.0014829999999999 |
| H | 0.5575053050000001 | 3.3137716629999998 | 0.0000000000000000 |
| H | 0.0000000000000000 | 4.4193718339999997 | 2.1437979999999999 |
| H | 0.5575053050000001 | 3.3137716629999998 | 4.2875959999999997 |
| H | 1.6808247000000001 | 1.1069077629999999 | 4.2861130000000003 |
| H | 2.2466158350000001 | 0.0072232569999997 | 2.1437979999999999 |
| C | 8.0182886870000001 | 3.4537443930000000 | 2.1437979999999999 |
| C | 8.3332679030000012 | 2.8318205740000000 | 0.9375929999999999 |
| C | 8.3332679030000012 | 2.8318205740000000 | 3.3500030000000001 |
| C | 8.9621537409999998 | 1.5879866770000000 | 0.9368439999999998 |
| C | 9.2749091900000007 | 0.9647566069999998 | 2.1437979999999999 |
| C | 8.9621537409999998 | 1.5879866770000000 | 3.3507520000000000 |
| H | 7.5176184829999997 | 4.4121485770000000 | 2.1437979999999999 |
| H | 8.0834096190000011 | 3.3124640710000000 | 4.2861130000000003 |
| H | 9.2067290130000004 | 1.1056001709999999 | 4.2875959999999997 |
| H | 9.7642343179999997 | 0.0000000000000000 | 2.1437979999999999 |
| H | 9.2067290130000004 | 1.1056001709999999 | 0.0000000000000000 |
| H | 8.0834096190000011 | 3.3124640710000000 | 0.0014829999999999 |

T 1: BE = 2.2 kcal/mol

| | | | |
|---|---|---|---|
| C | 3.8711150000000005 | 2.1435659999999999 | 2.4753989999999999 |
| C | 3.1724760000000005 | 0.9375579999999999 | 2.4753989999999999 |
| C | 1.7785590000000004 | 0.9363889999999999 | 2.4753989999999999 |
| C | 1.0815750000000004 | 2.1435659999999999 | 2.4753989999999999 |
| C | 1.7785590000000004 | 3.3507429999999996 | 2.4753989999999999 |
| C | 3.1724760000000005 | 3.3495739999999996 | 2.4753989999999999 |
| H | 4.9517300000000004 | 2.1435659999999999 | 2.4753989999999999 |
| H | 3.7157280000000004 | 0.0019269999999998 | 2.4753989999999999 |
| H | 1.2371550000000004 | 0.0000000000000000 | 2.4753989999999999 |
| H | 0.0000000000000000 | 2.1435659999999999 | 2.4753989999999999 |
| H | 1.2371550000000004 | 4.2871319999999997 | 2.4753989999999999 |
| H | 3.7157280000000004 | 4.2852049999999995 | 2.4753989999999999 |
| C | 7.1410130670000003 | 2.1435659999999999 | 1.0813359999999999 |
| C | 7.1414890670000002 | 3.3508040000000001 | 1.7783519999999999 |
| C | 7.1414890670000002 | 3.3508040000000001 | 3.1724459999999999 |
| C | 7.1410130670000003 | 2.1435659999999999 | 3.8694620000000000 |
| C | 7.1414890670000002 | 0.9363279999999998 | 3.1724459999999999 |
| C | 7.1414890670000002 | 0.9363279999999998 | 1.7783519999999999 |
| H | 7.1371830670000005 | 2.1435659999999999 | 0.0000000000000000 |
| H | 7.1405370670000003 | 4.2871310000000005 | 1.2371669999999999 |
| H | 7.1405370670000003 | 4.2871310000000005 | 3.7136309999999999 |
| H | 7.1371830670000005 | 2.1435659999999999 | 4.9507979999999998 |
| H | 7.1405370670000003 | 0.0000009999999997 | 3.7136309999999999 |
| H | 7.1405370670000003 | 0.0000009999999997 | 1.2371669999999999 |

T 2: BE = 2.8 kcal/mol

| | | | |
|---|---|---|---|
| C | 2.4753989999999999 | 2.1435659999999999 | 3.5136630000000002 |
| C | 2.4753989999999999 | 0.9375579999999999 | 4.2123020000000002 |
| C | 2.4753989999999999 | 0.9363889999999999 | 5.6062189999999994 |
| C | 2.4753989999999999 | 2.1435659999999999 | 6.3032029999999999 |
| C | 2.4753989999999999 | 3.3507429999999996 | 5.6062189999999994 |
| C | 2.4753989999999999 | 3.3495739999999996 | 4.2123020000000002 |
| H | 2.4753989999999999 | 2.1435659999999999 | 2.4330479999999999 |
| H | 2.4753989999999999 | 0.0019269999999998 | 3.6690499999999999 |
| H | 2.4753989999999999 | 0.0000000000000000 | 6.1476229999999994 |
| H | 2.4753989999999999 | 2.1435659999999999 | 7.3847780000000007 |
| H | 2.4753989999999999 | 4.2871319999999997 | 6.1476229999999994 |
| H | 2.4753989999999999 | 4.2852049999999995 | 3.6690499999999999 |
| C | 1.0813359999999999 | 2.1435659999999999 | 0.0004759999999999 |
| C | 1.7783519999999999 | 3.3508040000000001 | 0.0000000000000000 |
| C | 3.1724459999999999 | 3.3508040000000001 | 0.0000000000000000 |
| C | 3.8694620000000000 | 2.1435659999999999 | 0.0004759999999999 |
| C | 3.1724459999999999 | 0.9363279999999998 | 0.0000000000000000 |
| C | 1.7783519999999999 | 0.9363279999999998 | 0.0000000000000000 |
| H | 0.0000000000000000 | 2.1435659999999999 | 0.0043060000000001 |
| H | 1.2371669999999999 | 4.2871310000000005 | 0.0009519999999998 |
| H | 3.7136309999999999 | 4.2871310000000005 | 0.0009519999999998 |
| H | 4.9507979999999998 | 2.1435659999999999 | 0.0043060000000001 |
| H | 3.7136309999999999 | 0.0000009999999997 | 0.0009519999999998 |
| H | 1.2371669999999999 | 0.0000009999999997 | 0.0009519999999998 |

T 3: BE = 2.25 kcal/mol

| | | | |
|---|---|---|---|
| C | 3.8711150000000005 | 2.1435659999999999 | 2.4753989999999999 |
| C | 3.1724760000000005 | 0.9375579999999999 | 2.4753989999999999 |
| C | 1.7785590000000004 | 0.9363889999999999 | 2.4753989999999999 |
| C | 1.0815750000000004 | 2.1435659999999999 | 2.4753989999999999 |
| C | 1.7785590000000004 | 3.3507429999999996 | 2.4753989999999999 |
| C | 3.1724760000000005 | 3.3495739999999996 | 2.4753989999999999 |
| H | 4.9517300000000004 | 2.1435659999999999 | 2.4753989999999999 |
| H | 3.7157280000000004 | 0.0019269999999998 | 2.4753989999999999 |
| H | 1.2371550000000004 | 0.0000000000000000 | 2.4753989999999999 |
| H | 0.0000000000000000 | 2.1435659999999999 | 2.4753989999999999 |
| H | 1.2371550000000004 | 4.2871319999999997 | 2.4753989999999999 |
| H | 3.7157280000000004 | 4.2852049999999995 | 2.4753989999999999 |
| C | 7.8708798670000002 | 2.1435659999999999 | 1.0813359999999999 |
| C | 7.8713558670000001 | 3.3508040000000001 | 1.7783519999999999 |
| C | 7.8713558670000001 | 3.3508040000000001 | 3.1724459999999999 |
| C | 7.8708798670000002 | 2.1435659999999999 | 3.8694620000000000 |
| C | 7.8713558670000001 | 0.9363279999999998 | 3.1724459999999999 |
| C | 7.8713558670000001 | 0.9363279999999998 | 1.7783519999999999 |
| H | 7.8670498670000004 | 2.1435659999999999 | 0.0000000000000000 |
| H | 7.8704038670000003 | 4.2871310000000005 | 1.2371669999999999 |
| H | 7.8704038670000003 | 4.2871310000000005 | 3.7136309999999999 |
| H | 7.8670498670000004 | 2.1435659999999999 | 4.9507979999999998 |
| H | 7.8704038670000003 | 0.0000009999999997 | 3.7136309999999999 |
| H | 7.8704038670000003 | 0.0000009999999997 | 1.2371669999999999 |

T 4: BE = 1.12 kcal/mol

| | | | |
|---|---|---|---|
| C | 3.8711150000000005 | 2.1435659999999999 | 2.4753989999999999 |
| C | 3.1724760000000005 | 0.9375579999999999 | 2.4753989999999999 |
| C | 1.7785590000000004 | 0.9363889999999999 | 2.4753989999999999 |
| C | 1.0815750000000004 | 2.1435659999999999 | 2.4753989999999999 |
| C | 1.7785590000000004 | 3.3507429999999996 | 2.4753989999999999 |
| C | 3.1724760000000005 | 3.3495739999999996 | 2.4753989999999999 |
| H | 4.9517300000000004 | 2.1435659999999999 | 2.4753989999999999 |
| H | 3.7157280000000004 | 0.0019269999999998 | 2.4753989999999999 |
| H | 1.2371550000000004 | 0.0000000000000000 | 2.4753989999999999 |
| H | 0.0000000000000000 | 2.1435659999999999 | 2.4753989999999999 |
| H | 1.2371550000000004 | 4.2871319999999997 | 2.4753989999999999 |
| H | 3.7157280000000004 | 4.2852049999999995 | 2.4753989999999999 |
| C | 8.6007466670000010 | 2.1435659999999999 | 1.0813359999999999 |
| C | 8.6012226670000000 | 3.3508040000000001 | 1.7783519999999999 |
| C | 8.6012226670000000 | 3.3508040000000001 | 3.1724459999999999 |
| C | 8.6007466670000010 | 2.1435659999999999 | 3.8694620000000000 |
| C | 8.6012226670000000 | 0.9363279999999998 | 3.1724459999999999 |
| C | 8.6012226670000000 | 0.9363279999999998 | 1.7783519999999999 |
| H | 8.5969166670000003 | 2.1435659999999999 | 0.0000000000000000 |
| H | 8.6002706670000002 | 4.2871310000000005 | 1.2371669999999999 |
| H | 8.6002706670000002 | 4.2871310000000005 | 3.7136309999999999 |
| H | 8.5969166670000003 | 2.1435659999999999 | 4.9507979999999998 |
| H | 8.6002706670000002 | 0.0000009999999997 | 3.7136309999999999 |
| H | 8.6002706670000002 | 0.0000009999999997 | 1.2371669999999999 |

T 5: BE = 0.35 kcal/mol

| | | | |
|---|---|---|---|
| C | 3.8711150000000005 | 2.1435659999999999 | 2.4753989999999999 |
| C | 3.1724760000000005 | 0.9375579999999999 | 2.4753989999999999 |
| C | 1.7785590000000004 | 0.9363889999999999 | 2.4753989999999999 |
| C | 1.0815750000000004 | 2.1435659999999999 | 2.4753989999999999 |
| C | 1.7785590000000004 | 3.3507429999999996 | 2.4753989999999999 |
| C | 3.1724760000000005 | 3.3495739999999996 | 2.4753989999999999 |
| H | 4.9517300000000004 | 2.1435659999999999 | 2.4753989999999999 |
| H | 3.7157280000000004 | 0.0019269999999998 | 2.4753989999999999 |
| H | 1.2371550000000004 | 0.0000000000000000 | 2.4753989999999999 |
| H | 0.0000000000000000 | 2.1435659999999999 | 2.4753989999999999 |
| H | 1.2371550000000004 | 4.2871319999999997 | 2.4753989999999999 |
| H | 3.7157280000000004 | 4.2852049999999995 | 2.4753989999999999 |
| C | 9.8171913330000002 | 2.1435659999999999 | 1.0813359999999999 |
| C | 9.8176673329999993 | 3.3508040000000001 | 1.7783519999999999 |
| C | 9.8176673329999993 | 3.3508040000000001 | 3.1724459999999999 |
| C | 9.8171913330000002 | 2.1435659999999999 | 3.8694620000000000 |
| C | 9.8176673329999993 | 0.9363279999999998 | 3.1724459999999999 |
| C | 9.8176673329999993 | 0.9363279999999998 | 1.7783519999999999 |
| H | 9.8133613329999996 | 2.1435659999999999 | 0.0000000000000000 |
| H | 9.8167153330000012 | 4.2871310000000005 | 1.2371669999999999 |
| H | 9.8167153330000012 | 4.2871310000000005 | 3.7136309999999999 |
| H | 9.8133613329999996 | 2.1435659999999999 | 4.9507979999999998 |
| H | 9.8167153330000012 | 0.0000009999999997 | 3.7136309999999999 |
| H | 9.8167153330000012 | 0.0000009999999997 | 1.2371669999999999 |

Structures and binding energies of the ethane (E) dimer.

E 1: BE = -0.73 kcal/mol

| | | | |
|---|---|---|---|
| C | 0.0000000000000000 | -13.0744959999999999 | 1.5460059999999987 |
| C | 0.0000000000000000 | -13.0744959999999999 | 0.0000000000000000 |
| H | 0.0000000000000000 | -12.0499999999999989 | 1.9439980000000010 |
| H | -0.8872400000000003 | -13.5867439999999995 | 1.9439980000000010 |
| H | 0.8872399999999990 | -13.5867439999999995 | 1.9439980000000010 |
| H | 0.0000000000000000 | -14.0989920000000009 | -0.3979920000000011 |
| H | -0.8872400000000003 | -12.5622480000000003 | -0.3979920000000011 |
| H | 0.8872399999999990 | -12.5622480000000003 | -0.3979920000000011 |
| C | 0.0000000000000000 | -9.5744959999999999 | 1.5460059999999987 |
| C | 0.0000000000000000 | -9.5744959999999999 | 0.0000000000000000 |
| H | 0.0000000000000000 | -8.5500000000000025 | 1.9439980000000010 |
| H | -0.8872400000000003 | -10.0867440000000013 | 1.9439980000000010 |
| H | 0.8872399999999990 | -10.0867440000000013 | 1.9439980000000010 |
| H | 0.0000000000000000 | -10.5989920000000009 | -0.3979920000000011 |
| H | -0.8872400000000003 | -9.0622480000000003 | -0.3979920000000011 |
| H | 0.8872399999999990 | -9.0622480000000003 | -0.3979920000000011 |

E 2: BE = 0.988 kcal/mol

| | | | |
|---|---|---|---|
| C | 0.0000000000000000 | -13.0744959999999999 | 1.5460059999999987 |
| C | 0.0000000000000000 | -13.0744959999999999 | 0.0000000000000000 |
| H | 0.0000000000000000 | -12.0499999999999989 | 1.9439980000000010 |
| H | -0.8872400000000003 | -13.5867439999999995 | 1.9439980000000010 |
| H | 0.8872399999999990 | -13.5867439999999995 | 1.9439980000000010 |
| H | 0.0000000000000000 | -14.0989920000000009 | -0.3979920000000011 |
| H | -0.8872400000000003 | -12.5622480000000003 | -0.3979920000000011 |
| H | 0.8872399999999990 | -12.5622480000000003 | -0.3979920000000011 |
| C | 0.0000000000000000 | -9.1744959999999995 | 1.5460059999999987 |
| C | 0.0000000000000000 | -9.1744959999999995 | 0.0000000000000000 |
| H | 0.0000000000000000 | -8.1500000000000004 | 1.9439980000000010 |
| H | -0.8872400000000003 | -9.6867440000000009 | 1.9439980000000010 |
| H | 0.8872399999999990 | -9.6867440000000009 | 1.9439980000000010 |
| H | 0.0000000000000000 | -10.1989920000000023 | -0.3979920000000011 |
| H | -0.8872400000000003 | -8.6622480000000017 | -0.3979920000000011 |
| H | 0.8872399999999990 | -8.6622480000000017 | -0.3979920000000011 |

E 3: BE = 1.031 kcal/mol

| | | | |
|---|---|---|---|
| C | 0.0000000000000000 | -13.0744959999999999 | 1.5460059999999987 |
| C | 0.0000000000000000 | -13.0744959999999999 | 0.0000000000000000 |
| H | 0.0000000000000000 | -12.0499999999999989 | 1.9439980000000010 |
| H | -0.8872400000000003 | -13.5867439999999995 | 1.9439980000000010 |
| H | 0.8872399999999990 | -13.5867439999999995 | 1.9439980000000010 |
| H | 0.0000000000000000 | -14.0989920000000009 | -0.3979920000000011 |
| H | -0.8872400000000003 | -12.5622480000000003 | -0.3979920000000011 |
| H | 0.8872399999999990 | -12.5622480000000003 | -0.3979920000000011 |
| C | 0.0000000000000000 | -8.9744959999999985 | 1.5460059999999987 |
| C | 0.0000000000000000 | -8.9744959999999985 | 0.0000000000000000 |
| H | 0.0000000000000000 | -7.9500000000000020 | 1.9439980000000010 |
| H | -0.8872400000000003 | -9.4867440000000016 | 1.9439980000000010 |
| H | 0.8872399999999990 | -9.4867440000000016 | 1.9439980000000010 |
| H | 0.0000000000000000 | -9.9989920000000012 | -0.3979920000000011 |
| H | -0.8872400000000003 | -8.4622480000000024 | -0.3979920000000011 |
| H | 0.8872399999999990 | -8.4622480000000024 | -0.3979920000000011 |

E 4: BE = 0.922 kcal/mol

| | | | |
|---|---|---|---|
| C | 0.0000000000000000 | -13.0744959999999999 | 1.5460059999999987 |
| C | 0.0000000000000000 | -13.0744959999999999 | 0.0000000000000000 |
| H | 0.0000000000000000 | -12.0499999999999989 | 1.9439980000000010 |
| H | -0.8872400000000003 | -13.5867439999999995 | 1.9439980000000010 |
| H | 0.8872399999999990 | -13.5867439999999995 | 1.9439980000000010 |
| H | 0.0000000000000000 | -14.0989920000000009 | -0.3979920000000011 |
| H | -0.8872400000000003 | -12.5622480000000003 | -0.3979920000000011 |
| H | 0.8872399999999990 | -12.5622480000000003 | -0.3979920000000011 |
| C | 0.0000000000000000 | -8.7744959999999992 | 1.5460059999999987 |
| C | 0.0000000000000000 | -8.7744959999999992 | 0.0000000000000000 |
| H | 0.0000000000000000 | -7.7500000000000018 | 1.9439980000000010 |
| H | -0.8872400000000003 | -9.2867440000000023 | 1.9439980000000010 |
| H | 0.8872399999999990 | -9.2867440000000023 | 1.9439980000000010 |
| H | 0.0000000000000000 | -9.7989920000000019 | -0.3979920000000011 |
| H | -0.8872400000000003 | -8.2622480000000014 | -0.3979920000000011 |
| H | 0.8872399999999990 | -8.2622480000000014 | -0.3979920000000011 |

E 5: BE = 0.433 kcal/mol

| | | | |
|---|---|---|---|
| C | 0.0000000000000000 | -13.0744959999999999 | 1.5460059999999987 |
| C | 0.0000000000000000 | -13.0744959999999999 | 0.0000000000000000 |
| H | 0.0000000000000000 | -12.0499999999999989 | 1.9439980000000010 |
| H | -0.8872400000000003 | -13.5867439999999995 | 1.9439980000000010 |
| H | 0.8872399999999990 | -13.5867439999999995 | 1.9439980000000010 |
| H | 0.0000000000000000 | -14.0989920000000009 | -0.3979920000000011 |
| H | -0.8872400000000003 | -12.5622480000000003 | -0.3979920000000011 |
| H | 0.8872399999999990 | -12.5622480000000003 | -0.3979920000000011 |
| C | 0.0000000000000000 | -8.0744959999999999 | 1.5460059999999987 |
| C | 0.0000000000000000 | -8.0744959999999999 | 0.0000000000000000 |
| H | 0.0000000000000000 | -7.0500000000000025 | 1.9439980000000010 |
| H | -0.8872400000000003 | -8.5867440000000030 | 1.9439980000000010 |
| H | 0.8872399999999990 | -8.5867440000000030 | 1.9439980000000010 |
| H | 0.0000000000000000 | -9.0989920000000009 | -0.3979920000000011 |
| H | -0.8872400000000003 | -7.5622480000000021 | -0.3979920000000011 |
| H | 0.8872399999999990 | -7.5622480000000021 | -0.3979920000000011 |

E 6: BE =  0.005 kcal/mol

| | | | |
|---|---|---|---|
| C | 0.0000000000000000 | -13.0744959999999999 | 1.5460059999999987 |
| C | 0.0000000000000000 | -13.0744959999999999 | 0.0000000000000000 |
| H | 0.0000000000000000 | -12.0499999999999989 | 1.9439980000000010 |
| H | -0.8872400000000003 | -13.5867439999999995 | 1.9439980000000010 |
| H | 0.8872399999999990 | -13.5867439999999995 | 1.9439980000000010 |
| H | 0.0000000000000000 | -14.0989920000000009 | -0.3979920000000011 |
| H | -0.8872400000000003 | -12.5622480000000003 | -0.3979920000000011 |
| H | 0.8872399999999990 | -12.5622480000000003 | -0.3979920000000011 |
| C | 0.0000000000000000 | -3.0744960000000008 | 1.5460059999999987 |
| C | 0.0000000000000000 | -3.0744960000000008 | 0.0000000000000000 |
| H | 0.0000000000000000 | -2.0500000000000020 | 1.9439980000000010 |
| H | -0.8872400000000003 | -3.5867440000000022 | 1.9439980000000010 |
| H | 0.8872399999999990 | -3.5867440000000022 | 1.9439980000000010 |
| H | 0.0000000000000000 | -4.0989920000000009 | -0.3979920000000011 |
| H | -0.8872400000000003 | -2.5622480000000007 | -0.3979920000000011 |
| H | 0.8872399999999990 | -2.5622480000000007 | -0.3979920000000011 |

**Figure ESM1:** Sample Gaussian-09 (and earlier versions) input file demonstrating the use of carbon and hydrogen DCPs.

```
#B3LYP Gen Pseudo=Read OPT

Methane Dimer: Demonstration of DCP use

0 1
C      1.813080      -0.000001      -0.000019
H      1.448279      -0.878030      0.540132
H      2.906518      0.000026       -0.000092
H      1.448231      0.906738       0.490378
H      1.448112      -0.028731      -1.030295
C      -1.813079     0.000001       0.000019
H      -2.906513     -0.000468      -0.000296
H      -1.448453     0.878074       -0.540133
H      -1.447709     -0.906672      -0.490080
H      -1.448472     0.029059       1.030386

C H 0
6-31+G(2d,2p)
****

C 0
C 3 0
F and up
 3
  2  0.091556053  0.000025303
  2  0.044472350  0.000137829
  2  0.019075560 -0.000000056
S-F
 2
  2  0.075790561  0.000003145
  2  0.039119707  0.001009080
P-F
 2
  2  0.131194450 -0.000000531
  2  0.045246336 -0.003143976
D-F
 1
  2  0.033941983 -0.002000967
H 0
H 1 0
P an up
 3
  2   0.120883601    0.000231333
  2   0.044528578   -0.000070677
  2   0.005658790   -0.000000451
S-P
 1
  2   0.174740501   -0.000049845
```

**Table ESM1**. Signed errors in binding energies (SE) calculated for the non-covalently bonded dimers of the S22B benchmark set using B3LYP/6-31+G(2d,2p) with dispersion-correcting potentials without (Non-CP) counterpoise corrections. Overall performance statistics for the entire S22B set are provided at the bottom of the Table. All data are in kcal/mol except percent values.

| Dimer | Accepted High-Level Value[a] | SE[b] |
|---|---|---|
| 2-pyridoxine - 2-aminopyridine | 16.934 | -0.11 |
| adenine-thymine (hydrogen-bonded) | 16.660 | -0.54 |
| adenine-thymine (stacked) | 11.730 | -0.98 |
| benzene-water | 3.275 | 0.17 |
| benzene-hydrogen cyanide | 4.541 | 0.16 |
| benzene-indole (stacked) | 4.524 | 0.20 |
| benzene-indole (T-shaped) | 5.627 | -0.34 |
| benzene-methane | 1.448 | 0.17 |
| benzene-ammonia | 2.312 | 0.03 |
| ethylene dimer | 1.472 | -0.13 |
| ethylene-ethyne | 1.496 | 0.31 |
| formamide dimer | 16.062 | -0.17 |
| formic acid dimer | 18.753 | -0.07 |
| water dimer | 4.989 | 0.37 |
| methane dimer | 0.527 | -0.10 |
| ammonia dimer | 3.133 | 0.06 |
| phenol dimer | 7.097 | -0.46 |
| pyrazine dimer | 4.255 | -0.38 |
| slipped-parallel benzene dimer | 2.654 | 0.27 |
| T-shaped benzene dimer | 2.717 | 0.09 |
| uracil dimer (hydrogen bonded) | 20.641 | -0.45 |
| uracil dimer (stacked) | 9.805 | -0.34 |
| Mean Absolute/Signed Error | | 0.27/-0.10 |
| Mean Absolute/Signed % Error | | 6.30/0.08 |

[b]Defined as BE(DCP) – BE(Reference)

**Table ESM2**. Energies, in kcal/mol, associated with the processes in each of the indicated 10 sub-databases of the GMTKN30 database. Calculations were performed using B3LYP/6-31+G(2d,2p) without corrections for dispersion, with the DCPs provided in reference **Error! Bookmark not defined.**, with the DCPs described in the present work and with the D3 correction of reference **Error! Bookmark not defined.**.

BHPERI

| # | B3LYP | B3LYP-DCP (ref 19) | B3LYP-DCP (this work) | B3LYP-D3 | ref |
|---|---|---|---|---|---|
| 1 | 33.48 | 41.15 | 35.27 | 33.06 | 35.34 |
| 2 | 30.34 | 26.99 | 27.42 | 28.31 | 30.92 |
| 3 | 27.49 | 25.35 | 25.99 | 27.29 | 28.30 |
| 4 | 38.04 | 36.20 | 36.01 | 36.81 | 39.56 |
| 5 | 27.24 | 23.40 | 26.43 | 27.53 | 28.18 |
| 6 | 35.06 | 33.71 | 31.53 | 32.57 | 35.64 |
| 7 | 24.65 | 13.73 | 16.24 | 17.96 | 22.14 |
| 8 | 22.13 | 11.43 | 12.51 | 15.02 | 18.26 |
| 9 | 21.23 | 7.48 | 8.43 | 11.63 | 9.80 |
| 10 | 22.33 | 34.26 | 25.08 | 22.64 | 23.55 |
| 11 | 26.01 | 22.10 | 22.48 | 22.59 | 26.30 |
| 12 | 19.49 | 14.02 | 14.82 | 15.53 | 18.10 |
| 13 | 15.78 | 12.39 | 10.48 | 11.45 | 12.20 |
| 14 | 13.08 | 11.82 | 8.79 | 9.29 | 11.10 |
| 15 | 8.53 | 4.33 | 3.43 | 4.40 | 5.30 |
| 16 | 8.51 | 3.69 | 2.89 | 4.09 | 4.00 |
| 17 | 14.84 | 8.87 | 9.44 | 10.44 | 11.50 |
| 18 | 8.71 | 1.86 | 2.57 | 3.68 | 4.00 |
| 19 | 3.70 | -1.75 | -2.23 | -1.24 | -1.40 |
| 20 | 22.09 | 11.17 | 12.35 | 14.98 | 14.98 |
| 22 | 27.16 | 18.01 | 18.19 | 21.23 | 19.81 |
| 23 | 33.08 | 23.61 | 23.24 | 26.65 | 25.36 |

DARC

| # | B3LYP | B3LYP-DCP (ref 19) | B3LYP-DCP (this work) | B3LYP-D3 | ref |
|---|---|---|---|---|---|
| 1 | -37.70 | -65.93 | -51.24 | -43.19 | -43.80 |
| 2 | -57.86 | -85.39 | -70.91 | -62.73 | -59.30 |
| 3 | -19.47 | -43.47 | -33.92 | -25.66 | -30.00 |
| 4 | -26.86 | -48.35 | -40.57 | -32.40 | -33.10 |
| 5 | -26.93 | -53.17 | -42.57 | -33.92 | -36.50 |
| 6 | -42.13 | -66.65 | -57.34 | -48.60 | -48.20 |
| 7 | -0.62 | -20.07 | -15.36 | -9.51 | -14.40 |
| 8 | -2.96 | -21.65 | -16.92 | -11.05 | -16.20 |
| 9 | -3.18 | -23.24 | -18.26 | -12.29 | -17.20 |
| 10 | -5.84 | 34.26 | -20.14 | -14.11 | -19.20 |
| 11 | -17.69 | 22.10 | -34.77 | -27.33 | -31.60 |
| 12 | -18.41 | 14.02 | -35.24 | -27.61 | -32.10 |
| 13 | -19.91 | 12.39 | -37.33 | -29.80 | -34.10 |
| 14 | -20.59 | 11.82 | -37.72 | -30.01 | -34.40 |

BSR36

| # | B3LYP | B3LYP-DCP (ref 19) | B3LYP-DCP (this work) | B3LYP-D3 | ref |
|---|---|---|---|---|---|
| 1 | 3.11 | 8.34 | 10.74 | 8.52 | 9.81 |
| 2 | 4.62 | 8.09 | 10.40 | 8.68 | 9.66 |
| 3 | 4.16 | 9.79 | 12.62 | 9.86 | 11.37 |
| 4 | 3.54 | 7.58 | 9.73 | 8.00 | 9.04 |
| 5 | 5.07 | 7.40 | 9.28 | 8.05 | 8.71 |
| 6 | 6.31 | 9.29 | 11.60 | 10.08 | 10.91 |
| 7 | 5.44 | 11.12 | 14.17 | 11.65 | 13.05 |
| 8 | 5.83 | 9.98 | 12.72 | 10.72 | 11.89 |
| 9 | 5.18 | 11.52 | 14.82 | 11.81 | 13.53 |
| 10 | 3.17 | 9.61 | 12.38 | 9.92 | 11.43 |
| 11 | 2.66 | 10.75 | 14.25 | 10.79 | 12.97 |
| 12 | 3.22 | 11.12 | 13.99 | 10.96 | 12.77 |
| 13 | 4.70 | 9.39 | 12.01 | 10.02 | 11.23 |
| 14 | 3.24 | 8.28 | 10.97 | 9.03 | 10.16 |
| 15 | 0.40 | 11.87 | 16.44 | 11.74 | 15.05 |
| 16 | -0.59 | -1.47 | 2.16 | 0.65 | 2.38 |
| 17 | 6.15 | 8.87 | 11.48 | 9.38 | 10.67 |
| 18 | 1.78 | 2.41 | 6.59 | 4.41 | 6.35 |
| 19 | 8.45 | 12.65 | 16.03 | 13.24 | 14.88 |
| 20 | 2.68 | 6.35 | 11.52 | 8.04 | 10.65 |
| 21 | 3.94 | 5.97 | 10.84 | 8.01 | 10.11 |
| 22 | 1.48 | 4.92 | 9.56 | 6.61 | 9.13 |
| 23 | 3.83 | 6.50 | 11.16 | 8.31 | 10.48 |
| 24 | 3.67 | 5.57 | 10.59 | 7.77 | 9.82 |
| 25 | 6.50 | 13.94 | 20.71 | 15.78 | 19.30 |
| 26 | -3.35 | 5.91 | 11.39 | 7.34 | 9.67 |
| 27 | 4.06 | 9.49 | 16.38 | 12.32 | 15.19 |
| 28 | 14.64 | 22.39 | 27.85 | 23.22 | 26.05 |
| 29 | 11.08 | 20.16 | 25.25 | 20.64 | 23.45 |
| 30 | 13.08 | 23.66 | 29.89 | 24.69 | 27.96 |
| 31 | 13.44 | 21.57 | 27.00 | 22.87 | 25.41 |
| 32 | 14.56 | 23.36 | 30.14 | 23.76 | 27.56 |
| 33 | 20.34 | 34.20 | 42.91 | 34.37 | 39.56 |
| 34 | 11.71 | 26.50 | 34.74 | 26.82 | 32.42 |
| 35 | 27.05 | 46.17 | 56.69 | 46.05 | 51.44 |
| 36 | 24.44 | 39.60 | 50.11 | 40.23 | 47.06 |

ISO34

| # | B3LYP | B3LYP-DCP (ref 19) | B3LYP-DCP (this work) | B3LYP-D3 | ref |
|---|---|---|---|---|---|
| 1 | -2.47 | -1.80 | -2.82 | -2.59 | 1.62 |
| 2 | 22.73 | 14.99 | 20.12 | 22.56 | 21.88 |
| 3 | 8.88 | -0.45 | 6.64 | 9.03 | 7.20 |
| 4 | 1.34 | 0.63 | 1.03 | 1.07 | 0.99 |
| 5 | 0.48 | 1.14 | 1.51 | 1.04 | 0.93 |
| 6 | 3.18 | 3.86 | 3.07 | 2.92 | 2.62 |
| 7 | 14.03 | 4.18 | 10.28 | 13.92 | 11.15 |
| 8 | 21.00 | 22.00 | 23.35 | 22.04 | 22.90 |
| 9 | 8.27 | 9.10 | 8.15 | 7.98 | 6.94 |
| 10 | 1.32 | 3.21 | 4.55 | 3.12 | 3.58 |
| 11 | -6.80 | 0.19 | 2.33 | -0.26 | 1.91 |
| 12 | 55.80 | 51.56 | 50.97 | 54.73 | 46.95 |
| 13 | 39.57 | 41.28 | 40.11 | 38.77 | 36.04 |
| 14 | 22.46 | 35.37 | 25.92 | 22.29 | 24.20 |
| 15 | 7.25 | 8.74 | 8.25 | 7.36 | 7.26 |
| 16 | 12.77 | 5.87 | 10.74 | 12.85 | 10.81 |
| 17 | 25.83 | 30.77 | 28.99 | 26.46 | 26.98 |
| 18 | 11.74 | 10.43 | 11.01 | 12.03 | 11.16 |
| 19 | 4.05 | 3.95 | 3.73 | 4.04 | 4.60 |
| 20 | 18.66 | 20.43 | 19.08 | 18.61 | 20.23 |
| 21 | 1.15 | 0.66 | 0.80 | 1.02 | 0.94 |
| 22 | 2.42 | 4.71 | 3.89 | 2.83 | 3.23 |
| 23 | 4.56 | 6.84 | 6.17 | 4.80 | 5.26 |
| 24 | 11.03 | 14.20 | 13.32 | 11.25 | 12.52 |
| 25 | 29.11 | 26.06 | 29.19 | 29.37 | 26.49 |
| 26 | 16.55 | 20.02 | 18.85 | 16.65 | 18.16 |
| 27 | 62.17 | 70.75 | 68.75 | 63.14 | 64.17 |
| 28 | 33.70 | 30.86 | 33.56 | 33.80 | 31.22 |
| 29 | 10.64 | 17.92 | 14.66 | 11.08 | 11.90 |
| 30 | 9.63 | 10.46 | 9.90 | 9.24 | 9.50 |
| 31 | 12.64 | 22.19 | 17.58 | 13.91 | 14.05 |
| 32 | 3.10 | 10.81 | 3.58 | 3.77 | 7.10 |
| 33 | 10.13 | 12.68 | 10.17 | 8.75 | 5.62 |
| 34 | 6.85 | 9.72 | 7.09 | 7.10 | 7.26 |

ISOL22

| # | B3LYP | B3LYP-DCP (ref 19) | B3LYP-DCP (this work) | B3LYP-D3 | ref |
|---|---|---|---|---|---|
| 1 | 21.10 | 51.75 | 41.17 | 32.56 | 40.59 |
| 2 | 7.98 | 11.80 | 8.52 | 7.85 | 11.68 |
| 3 | 25.00 | 49.94 | 36.97 | 29.24 | 34.94 |
| 4 | 22.83 | 33.04 | 24.16 | 22.67 | 25.89 |
| 5 | 8.36 | 24.37 | 18.99 | 14.29 | 18.79 |
| 6 | 47.15 | 15.88 | 27.91 | 36.54 | 18.30 |
| 7 | 18.98 | 25.53 | 23.82 | 22.11 | 22.31 |
| 8 | 3.37 | 10.17 | 8.57 | 5.40 | 7.91 |
| 9 | 35.62 | 36.59 | 34.95 | 36.01 | 38.13 |
| 11 | 30.57 | 38.37 | 35.40 | 31.88 | 35.08 |
| 12 | 5.15 | 9.09 | 9.97 | 6.93 | 5.20 |
| 13 | 2.79 | -3.56 | 7.70 | 5.31 | 3.87 |
| 14 | 23.03 | 27.57 | 22.46 | 22.53 | 22.59 |
| 15 | 0.30 | 19.46 | 9.28 | 1.79 | 11.07 |

PCONF

| # | B3LYP | B3LYP-DCP (ref 19) | B3LYP-DCP (this work) | B3LYP-D3 | ref |
|---|---|---|---|---|---|
| 1 | -4.21 | -0.18 | -0.05 | -0.56 | 0.14 |
| 2 | -1.07 | 0.43 | 0.72 | 0.32 | 0.90 |
| 3 | -4.58 | 0.60 | 0.47 | 1.74 | 1.15 |
| 4 | -2.60 | 0.81 | 0.93 | 0.28 | 0.79 |
| 5 | -3.83 | 0.99 | 0.80 | 1.87 | 1.31 |
| 6 | 0.02 | 0.78 | 1.50 | 1.23 | 1.87 |
| 7 | 1.02 | 1.75 | 1.85 | 2.03 | 2.37 |
| 8 | -3.26 | 1.46 | 1.38 | 2.38 | 2.07 |
| 9 | -1.70 | 0.99 | 1.72 | 2.09 | 2.51 |
| 10 | -1.71 | 1.12 | 1.51 | 0.98 | 2.04 |

ACONF

| # | B3LYP | B3LYP-DCP (ref 19) | B3LYP-DCP (this work) | B3LYP-D3 | ref |
|---|---|---|---|---|---|
| 1 | 1.01 | 0.58 | 0.64 | 0.67 | 0.60 |
| 2 | 1.03 | 0.60 | 0.64 | 0.65 | 0.61 |
| 3 | 2.03 | 0.67 | 0.96 | 1.04 | 0.96 |
| 4 | 3.72 | 3.09 | 2.92 | 2.82 | 2.81 |
| 5 | 1.01 | 0.55 | 0.60 | 0.61 | 0.60 |
| 6 | 1.06 | 0.60 | 0.64 | 0.65 | 0.60 |
| 7 | 2.16 | 0.71 | 0.98 | 1.06 | 0.93 |
| 8 | 2.10 | 1.15 | 1.24 | 1.28 | 1.18 |
| 9 | 2.08 | 1.22 | 1.31 | 1.33 | 1.30 |
| 10 | 3.22 | 0.93 | 1.39 | 1.51 | 1.25 |
| 11 | 3.70 | 2.90 | 2.74 | 2.67 | 2.63 |
| 12 | 3.70 | 3.00 | 2.81 | 2.71 | 2.74 |
| 13 | 4.66 | 3.61 | 3.40 | 3.33 | 3.28 |
| 14 | 4.79 | 3.13 | 3.14 | 3.03 | 3.08 |
| 15 | 6.59 | 5.28 | 5.05 | 4.81 | 4.93 |

IDISP

| # | B3LYP | B3LYP-DCP (ref 19) | B3LYP-DCP (this work) | B3LYP-D3 | ref |
|---|---|---|---|---|---|
| 1 | 19.64 | -15.65 | -7.72 | -2.71 | -9.00 |
| 2 | -75.10 | -54.80 | -55.33 | -59.57 | -58.50 |
| 3 | 7.26 | -0.61 | -2.45 | 0.48 | -1.90 |
| 4 | 24.89 | 11.17 | 8.17 | 11.74 | 8.20 |
| 5 | -7.56 | -4.02 | -3.86 | -3.48 | -3.10 |
| 6 | -17.87 | -1.99 | -2.52 | -1.11 | 0.40 |

ADIM6

| # | B3LYP | B3LYP-DCP (ref 19) | B3LYP-DCP (this work) | B3LYP-D3 | ref |
|---|---|---|---|---|---|
| 1 | -0.59 | 1.18 | 1.22 | 1.36 | 1.30 |
| 2 | -0.88 | 1.87 | 1.92 | 2.12 | 1.97 |
| 3 | -1.24 | 2.74 | 2.81 | 3.17 | 2.79 |
| 4 | -1.62 | 3.54 | 3.63 | 4.12 | 3.68 |
| 5 | -2.03 | 4.45 | 4.54 | 5.18 | 4.61 |
| 6 | -2.53 | 5.31 | 5.34 | 6.07 | 5.60 |